# A Study of Pseudo-Central Collision Events Observed in Pion-Nucleus Interactions

Tufail Ahmad[*] and Nazeer Ahmad

*Department of Physics, Aligarh Muslim University, Aligarh-202 002, India.*

The present study is based on the interactions caused by 340-GeV negative pions with emulsion nuclei. The main aim of this paper is to investigate some aspects of central collision events. Thus, the events in which the total number of charged shower particles is greater than or equal to twenty eight ($N_s \geq 28$) were chosen for the analysis. They are not exactly central collision events, but may be considered as pseudo-central collision events. The angular characteristics of relativistic charged secondaries have been studied in terms of pseudo-rapidity, and bimodality is found to be absent in the distributions. The mean pseudo-rapidity seems to be independent of grey and heavy particle multiplicities, which indicates its independence with number of collisions. Finally, the correlation between different particle multiplicities in this paper is discussed.

PACS numbers: 13.85, 25.80. Hp

Keywords: Relativistic charged particles, Grey, Black and heavily ionizing particle multiplicity, Pseudo-rapidity

[*]E-mail: tufailahmadphys@gmail.com

# I. INTRODUCTION

High-energy hadron-nucleus (hA) and nucleus-nucleus (AA) collisions have been investigated extensively by various workers [1-14] in the past. AA collisions may be explained as the superposition of many hA interactions which requires a detailed description of hadronic collisions. The interaction between elementary particles and nuclei is very important in order to get some insight into multiparticle production dynamics. The investigation of hA collisions is fundamental for studying the nature of the particles produced in such collisions. In high-energy multiparticle production, most of the work done is for inclusive interactions that are for all impact parameters, but fewer attempts have been made to study collisions with fewer impact parameters. Keeping this fact in mind, we chose to study events other than normal events. Therefore, we analyzed events beyond two standard deviations (σ) from the average value. In the case of inclusive data, the value of σ is equal to 6.48 ± 0.54 [7], and the mean value of relativistic charged particles, $<N_s>$, is 14.18 ± 0.08 [7]. In this way, $N_s \geq <N_s> + 2\sigma$, that is $N_s \geq 28$. Thus, events with $N_s \geq 28$ have been considered for the present study. We can say that these events are high multiplicity events. We cannot say that these events are exact central collision events, but may be considered as pseudo-central collision events.

We have analyzed the pseudo-central collision events observed in pion-nucleus collisions at the highest momentum of 340-GeV/c negative pions. For this purpose, the nuclear emulsion technique has been used. Nuclear emulsion is a detector that detects charged particles only. This technique for studying nuclear interactions has an advantage because it has high resolution. When a beam of high-energy particles is incident on an emulsion, secondary charged particles are produced, which is known as multiparticle production. The secondary particles leave behind tracks in the emulsion. Such tracks can be categorized as slow or relativistic charged particles. The slow particles are nothing, but target protons.

The results discussed are the rapidity distributions for inclusive data and pseudo-central events. The variation of the normalized pseudo-rapidity density with the pseudo-rapidity has been studied. The dependence of the mean pseudo-rapidity on the grey and the heavy ionizing particle multiplicity is presented. The correlations between different particle multiplicities have also been investigated.

# II. EXPERIMENTAL DETAILS

For the present investigation, the data were collected using several Ilford G5 emulsion pellicles exposed to a negative pion beam of 340-GeV/c momentum at the European Organization for Nuclear Research (CERN) Super proton Synchrotron (SPS). An interaction in the emulsion is called a star because of its very characteristic look. All the measurements were made under a 100X oil immersion objective on a compound microscope.

The events produced within 35 μm from the top or the bottom surfaces of the emulsion plates/pellicles were excluded from the final data. In order to ensure that the chosen data sample did not include any secondary interaction, we followed the primaries of each and every interaction back to the edge. Only those interactions whose primary remained almost parallel to the direction of the main beam were considered for study.

The tracks of produced particles were classified according to the usual emulsion terminology. The secondary particles leave tracks of their own, and they were classified on the basis of their specific ionization $g^*(=g/g_0)$, where g is the ionization of the track and $g_0$ is the ionization of the primary. Tracks with $g^* < 1.4$, $1.4 \leq g^* \leq 10$ and $g^* > 10$ were taken as shower, grey, and black tracks respectively. The numbers of shower, grey, and black tracks in an event are denoted by $N_s$, $N_g$ and $N_b$, respectively. When grey and black tracks are taken together in an event, they are called heavy tracks, $N_h$ (= $N_g$ + $N_b$). Other relevant details may be found in our earlier publications [5-8].

## III. RESULTS AND DISCUSSION

For the study of the angular characteristics of shower particles, we define the pseudo-rapidity variable, $\eta$, as

$$\eta = -\ln(\tan\theta/2), \quad (1)$$

where $\theta$ is the emission angle of the secondary particle. The normalized single-particle pseudo-rapidity distribution of charged secondaries is shown in Fig. 1 for the events with $N_s \geq 28$. In order to make a comparison for inclusive data, we also give the distribution for $N_h \geq 0$ in the same figure. The distributions do not show any bimodality (double bump structure). Similar results have been reported [2] for proton-nucleus collisions. Weak evidence of bimodality has been shown by Boos *et al.* [15]. Evidence of a bimodal structure supports the models of repeated intranuclear collisions. Because of the absence of a double bump structure, the present work is in disagreement with the intranuclear cascade model [16]. Kaur *et.al.* [17] and Babecki *et al.* [18] at 50- and 300-GeV/c pion-nucleus collisions reported similar results in their papers. However, some reported results [1,19] favor the intranuclear cascade model. Therefore, we cannot say firmly whether a bimodal structure in the pseudo-rapidity distribution will be observed or not.

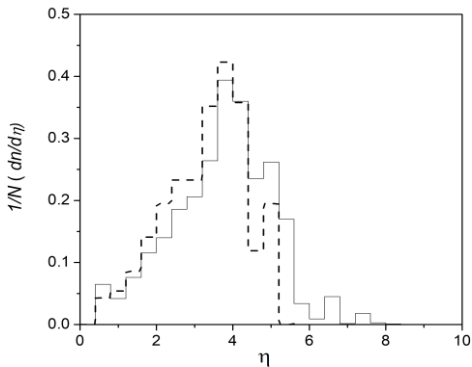
Fig. 1

To investigate the angular characteristics of charged secondaries in more detail, we define the normalized pseudo-rapidity density, $R(\eta)$, in the following way:

$$R(\eta) = \rho_{\pi^- A}(\eta)/\rho_{\pi^- N}(\eta), \quad (2)$$

where

$$\rho(\eta) = (1/N)(dN_s/d\eta) \quad (3)$$

are the shower particle densities for pion-nucleus and pion-nucleon interactions. These have been calculated for N pion interactions. The normalized pseudo-rapidity density distribution is displayed in Fig. 2. The variations of $R(\eta)$ with $\eta$ are very similar for the two data sets, i.e., for pseudo-central collisions and inclusive data. In the projectile fragmentation region, i.e., $\eta > 6$, the values of $R(\eta)$ have been reported [3] to be less than unity for proton-nucleus collision data. However, in our case, that is, for pion-nucleus collisions, we observe that $R(\eta)$ is less than unity for $\eta \geq 3.5$.

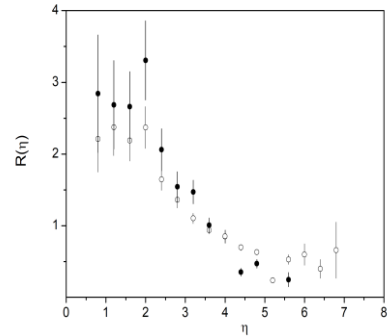
Fig. 2

Figure 3 shows the variation of the mean pseudo-rapidity, $\langle\eta\rangle$, with the number of grey and heavy ionizing particles. In one of our earlier papers [8] on inclusive data for pion-nucleus collisions, we reported that the mean pseudo-rapidity changes its form after some value of the grey or the heavy particle multiplicity. However, we did not observe any such trend for pseudo-central collision data in this research.

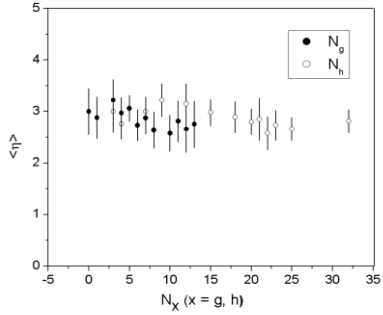

Fig. 3

The following equations were obtained after performing least-square fits:

$\langle\eta\rangle = (-0.01 \pm 0.00)N_h + (3.05 \pm 0.09)$, (4)

$\langle\eta\rangle = (-0.03 \pm 0.01)N_g + (3.08 \pm 0.08)$. (5)

The variation of mean pseudo-rapidity with the square root of the grey or the heavy particle multiplicity is shown in Fig. 4. The data is best represented by the following lines:

$\langle\eta\rangle = (-0.08 \pm 0.04)\sqrt{N_h} + (3.17 \pm 0.15)$, (6)

$\langle\eta\rangle = (-0.13 \pm 0.05)\sqrt{N_g} + (3.16 \pm 0.13)$. (7)

Thus, from eqs. (4) – (7), we conclude that the mean pseudo-rapidity does not vary very much with the mean number of intranuclear collisions, $\langle\nu\rangle$, made by the incident hadron because the grey and the heavy particle multiplicities are regarded as measures of the numbers of collisions.

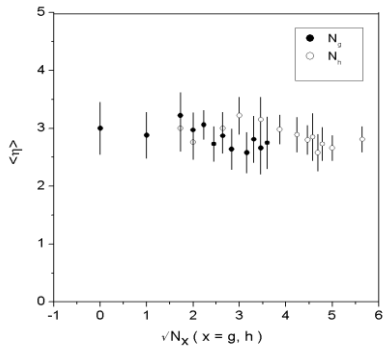

Fig. 4

In order to study the correlation between different particle multiplicities, we plot the variations of $\langle N_s\rangle$, $\langle N_b\rangle$ and $\langle N_h\rangle$ with different number of grey particles in Fig. 5. From the figure, we note that the increase in the values of $\langle N_b\rangle$ and $\langle N_h\rangle$ are almost linear, but in the case of the variation of $\langle N_b\rangle$ with $N_g$, a saturation effect is observed beyond $N_g = 6$, which is why a least-squares fit was performed up to $N_g = 6$ only, the following equations were obtained.

$\langle N_b\rangle = (1.39 \pm 0.35)N_g + (2.02 \pm 1.55)$, (8)

$\langle N_h\rangle = (2.16 \pm 0.22)N_g + (3.24 \pm 1.23)$. (9)

A similar result was observed for inclusive data [20]. As far as the variation in $\langle N_s\rangle$ with $N_g$ is concerned, no such behavior is seen.

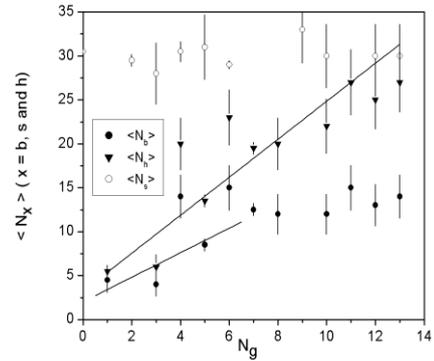

Fig. 5

### IV. CONCLUDING REMARKS

The conclusions that can be drawn from the study of pseudo-central collision events caused by 340-GeV negative pions are as follows:

(i) The double bump structure is found to be completely absent in the pseudo-rapidity distribution.
(ii) The normalized shower-particle pseudo-rapidity density, $R(\eta)$, is observed to be very similar for inclusive data and pseudo-central collision events.

(iii) The mean pseudo-rapidity shows almost no dependence on the number of collisions.

(iv) Almost linear increase in the variations of $<N_b>$ and $<N_h>$ with number of grey particles are observed; however, $<N_b>$ is found to become saturated beyond $N_g = 6$.

**Figure Caption**

Fig. 1. Normalized single-particle pseudo-rapidity distribution of charged shower particles for events with $N_s \geq 28$ (---) and $N_h \geq 0$ (–).

Fig. 2. Variation of the normalized pseudo-rapidity density, $R(\eta)$, with $\eta$ for pseudo-central collision events (●) and inclusive data (o).

Fig. 3. Variation of the mean pseudo-rapidity with the number of grey and heavily ionizing particles.

Fig. 4. Variation of mean the pseudo-rapidity with the square root of the grey and the heavily ionizing particle multiplicities.

Fig. 5. Dependence of the mean numbers of black, heavy and shower particles on the grey particle multiplicity.